\def\ket#1{\left|#1\right\rangle}

\def\tr{\mathop{\rm Tr}}
\makeatletter
\def\Hy@safe@activestrue{}
\makeatother
\documentclass[aps,prb,twocolumn,groupedaddress]{revtex4-1}
\usepackage[hypertex]{hyperref}
\usepackage{amsmath}
\usepackage{amsfonts}
\usepackage{amssymb}
\usepackage{graphicx}
\usepackage{epsfig}
\usepackage{textcomp}

\begin{document}

\title{Suppression of Hyperfine Dephasing by Spatial Exchange of Double
  Quantum Dots}
 \author{David Drummond}
\affiliation{University of California, Riverside, California, 92521, USA}
\author{Leonid P. Pryadko}
\affiliation{University of California, Riverside, California, 92521, USA}
\author{Kirill Shtengel}
\affiliation{University of California, Riverside, California, 92521, USA}
\date{\today}

\begin{abstract}
  We examine the logical qubit system of a pair of electron spins in double
  quantum dots.  Each electron experiences a different hyperfine interaction
  with the local nuclei of the lattice, leading to a relative phase
  difference, and thus decoherence.  Methods such as nuclei polarization,
  state narrowing, and spin-echo pulses have been proposed to delay
  decoherence.  Instead we propose to suppress hyperfine dephasing by
  adiabatic rotation of the dots in real space, leading to the same average
  hyperfine interaction.  We show that the additional effects due to the
  motion in the presence of spin-orbit coupling are still smaller than the hyperfine
  interaction, and result in an infidelity below $10^{-4}$ after ten decoupling cycles.  We discuss a possible experimental setup and
  physical constraints for this proposal.
\end{abstract}
\maketitle

\section{Introduction}
In recent years there has been great interest in the prospect of using
scalable solid state devices to implement quantum two-level systems(qubits)
for potential applications such as quantum computation.  One promising
candidate for a qubit is a pair of electron spins in quantum dots, which forms
a fault-tolerant subspace that is immune to collective
decoherence.\cite{levy-2002} However, each electron is still subject to the
local hyperfine interaction from the nuclear spins of the lattice, which leads
to dephasing of the individual electron spins.\cite{burkardloss-1999} There
have been several proposals to suppress this dephasing such as nuclear
polarization,\cite{burkardloss-1999, khaetskii-2003, gullans-2010} state
narrowing,\cite{klauser-2006} and spin-echo pulse
correction.\cite{khodjasteh-Lidar-2005, zhang-2007} While improved coherence
has been experimentally demonstrated using these techniques, the coherence
times desired for applications have proven very difficult to achieve.  For
example, pumping methods have been used to partially polarize nuclei, but the
nearly full polarization needed has yet to be achieved.\cite{klauser-2006,
  imamoglu-2003, Taylor-2003, reilly-2008}

Promising results have been shown using spin-echo sequences through the
exchange interaction between two spins.\cite{petta-2005, barthel-2010,
  vanWeperen-2011} The exchange is controlled by lowering the tunneling
barrier between the two quantum dots using quickly-controlled electric gates.
This leads to Rabi oscillations; a single $\pi$-pulse corresponds to
exchanging the two spins.  Sequences of such pulses effectively couple both spins to the same average hyperfine interaction resulting in improved
coherence times.  While several echo sequences can be performed using this
exchange, it is likely that the electrical gate noise and spatial variations in the Overhauser field remain the dominant sources of
dephasing.\cite{barthel-2010}

\begin{figure}[tbp]
 \begin{center}
   \includegraphics[width=0.38\columnwidth]{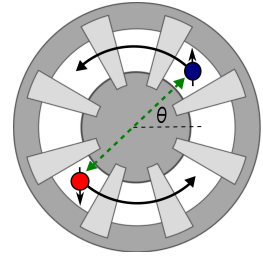}
   \caption{Suggested electrode geometry for a rotating double-quantum-dot
     qubit with top and bottom gates in different shades of gray.  Exchange
     gates via real space rotation, as opposed to tunneling, are expected to
     strongly reduce the qubit sensitivity to charge noise.
   }
  \label{fig:setup}
  \end{center}
\end{figure}

Rather than relying on interdot tunneling, we propose using a spatial exchange of the two quantum dots, allowing the two electrons to traverse the same path, spending the same time coupled to the local nuclei, as shown in Figure~\ref{fig:setup}.
Compared to exchange via tunneling, ideally, this should eliminate the effect
of the electrical gate noise.  On the other hand, in the presence of spin-orbit
coupling, such a motion of electrons may introduce additional errors.  In
this manuscript we analyze how these errors depend on the parameters of the motion and discuss the constraints and possible parameters for potential
implementation of this proposal.

\begin{figure}[btp]
  \centering
  \includegraphics[width = 0.8\columnwidth]{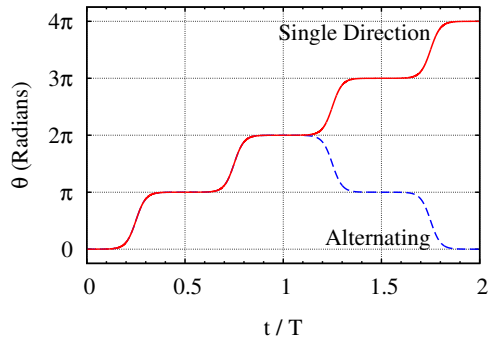}
  \caption{Time dependent adiabatic trajectories used in the simulations.
    Plotted is the position of the first dot parametrized in terms of the
    angle $\theta$ as a function of time $t$. The angle-dependent positions
    were defined as a sum of properly scaled and shifted hyperbolic tangents.
    Single direction rotations suppress the effect of a static Overhauser
    field but not of a time-varying one.  Longer rotation sequences like
    alternating forward-and-back suppress the effect of a linear in time
    Overhauser field.  Arbitary rotations on the Bloch sphere may be
    accomplished using additional operations when the dot is stationary,
    corresponding to the flat segments in the sequence.}
   \label{fig:protocols}
\end{figure}

We note that a similar setup with spatial exchange of
electrostatically-defined quantum dots has been discussed in relation
to holonomic quantum computation.\cite{golovach-2010,
  shitade-2010} However, the corresponding coherence estimates have
been done in the absence of a magnetic field.  Here we focus on the
parameter range characteristic of double-quantum-dot qubits and take into account typical magnetic fields of order $\agt 0.1$\,T.

The outline of the paper is as follows.  We define the Hamiltonian of
the double-quantum-dot qubit with spatial exchange in
Sec.~\ref{sec:setup}, derive the effective spin-only Hamiltonain for a
single spin in a moving quantum dot in Sec.~\ref{sec:transformations},
and calculate the single-qubit fidelity associated with a sequence of
double-dot rotations in Sec.~\ref{sec:analysis}.  We discuss simulations of a possible protocol in Sec.~\ref{sec:simulations}, and the constraints and corresponding characteristic time and
distance scales in Sec.~\ref{sec:experimental-proposal}, and conclude
with Sec.~\ref{sec:conclusion}.

\section{Setup}
\label{sec:setup}
We envision a qubit formed by a pair of quantum dots electrostatically defined
in a III-V semiconductor (e.g., GaAs/AlGaAs) heterostructure using a system of
top and bottom gates similar to that illustrated in Fig.~\ref{fig:setup}.  We
take the parameters of the dots to be similar to the experimental ones.~
\cite{petta-2005, barthel-2010, vanWeperen-2011}  Specifically, each dot
contains a single electron, with a typical dot-size quantization energy $\hbar
\omega_{d}\sim 1$meV.  The qubit is defined as the subspace of the singlet and
$m_{s}=0$ triplet states of the two electrons.  The triplet degeneracy is
removed by a uniform, constant magnetic field $B_0$ of at least $\sim 0.1$T
applied perpendicular to the sample, which creates a Zeeman gap of $\Delta
\sim 2.5$~{\textmu}eV.  The electrons interact with $N\sim10^6$ spins of the
lattice nuclei, leading to a different local hyperfine interaction for each
electron.  We will approximate this as a Zeeman interaction with a random,
fluctuating, non-uniform magnetic ``Overhauser'' field
$B_\text{N}\sim 1$mT.

To prevent dephasing, both quantum dots will be moved along the same
trajectory in a time much shorter than the relaxation time of the nuclear
spins, $t_\text{nuc}\sim100$~{\textmu}s, so we can assume that the Overhauser
field is quasi-static.  In order to reduce the sensitivity to charge noise the
distance between the dots must be significantly greater than the size of each
quantum dot, $a\sim100$nm.  Due to this spatial separation between the dots, we can treat the Hamiltonian of each electron separately,
\begin{equation}
H=H_\text{d}(\mathbf{r}_0)+H_\text{Z}
+V_\text{Z}(\mathbf{r},t)+H_\text{SO}+V(\mathbf{r},t).
\end{equation}
Here the dot Hamiltonian is given by
\begin{equation}
  \label{eq:ham-dot}
    H_\text{d}(\mathbf{r}_0)=\frac{p^2}{2m}+U(\mathbf{r}-\mathbf{r}_0(t)),
\end{equation}
with the canonical momentum $\mathbf{p}=\mathbf{P}+e\mathbf{A}/c$ and the
confining potential $U$ centered at $\mathbf{r}_0\equiv \mathbf{r}_0(t)$; the
Zeeman Hamiltonians for the externally-applied and Overhauser fields are
respectively
\begin{align}
  H_\text{Z}&=\frac{g \mu_\text{B}}{2} \mathbf{B}_0\cdot \boldsymbol \sigma,\\
  V_\text{Z}(\mathbf{r},t)&=\frac{g \mu_\text{B}}{2}
  \mathbf{B}_\text{N}(\mathbf r,t)\cdot \boldsymbol \sigma,
\end{align}
and the spin-orbit Hamiltonian is given by
\begin{equation}
  H_\text{SO}=\sigma^i C^{ij}p^j\label{eq:ham-so}.
\end{equation}
The last term, $V\biglb(\mathbf{r},t\bigrb)$, accounts for additional effects originating from
disorder, variation of the dot potential as it moves due to imperfections of
the confining potential, as well as phonons.  The spin orbit coupling coefficients $C_{ij}$ in
Eq.~(\ref{eq:ham-so}) incorporate both Dresselhaus (originating from the lack of the inversion symmetry of the lattice) with $C^{yy}=-C^{xx}=\beta$, and Rashba terms (structural
inversion asymmetry due to the quantum well) with $C^{xy}=-C^{yx}=\alpha$.  This specific form 
assumes that the growth of the semiconductor heterostructure and the quantum
well asymmetry are in the positive $z$ direction,\cite{Silsbee-2004,
  Zawadzki-2004} so all the matrix elements that involve $z$ are zero.

For numerical estimates we will use the effective electron mass
$m\sim0.067m_\text{e}$, and assume $\alpha \simeq \beta$ with values ranging
from $10^3$ to $10^4$ m/s, as appropriate for typical GaAs
heterostructures.\cite{Silsbee-2004} Based on the above values, we have
$m\beta^2 \ll g\mu_\text{B} B \ll \hbar \omega_\text{d}$, so we will ignore
terms quadratic in the spin-orbit coupling for the following analysis.

\section{Effective single-dot Hamiltonian}
\label{sec:transformations}
We now go into the moving reference frame of the dot using the translation
operator
\begin{align}
  \Psi(t) &\to e^{-\frac{i}{\hbar}\mathbf{P} \cdot \mathbf r_o(t)}\Psi(t), \\
  H &\to e^{\frac{i}{\hbar}\mathbf{P} \cdot \mathbf r_o(t)} H
  e^{-\frac{i}{\hbar}\mathbf{P} \cdot \mathbf r_o(t)} - \mathbf v_0(t) \cdot
  \mathbf{P},
\label{eq:translation}
\end{align}
which introduces the additional term proportional to the dot's velocity,
$\mathbf{v}_0\equiv \dot {\mathbf{r}}_0$, and removes the $\mathbf r_0$ from
the confining potential, $U(\mathbf{r}-\mathbf{r}_0) \to U(\mathbf{r})$.  This
also affects the vector potential, $\mathbf{A(r) \to A(r+r}_0)$, in the dot
Hamiltonian and spin-orbit term, which can be reversed with an appropriate
gauge transformation.  We select the symmetric gauge, $\mathbf A(\mathbf{r} +
\mathbf{r}_0)=\frac{1}{2}\mathbf B\times (\mathbf r+ \mathbf r_0) $, and
transform $\mathbf A \to \mathbf A + \mathbf \nabla f$ with $f=-\frac{1}{2}
\mathbf{r} \cdot (\mathbf B\times \mathbf r_0)$ which results in
\begin{align}
\Psi &\to \Psi \, \exp(-\frac{ie}{\hbar c}f), \label{psitran}\\
\mathbf P &\to \mathbf P-\frac{e}{c}\mathbf \nabla f. \label{ptran}
\end{align}
These transformations introduce two additional terms in the time-dependent
Schr\"{o}dinger equation $i\hbar\partial\Psi/\partial t = H\Psi$. The first
term, $-({e}/{2c}) \mathbf{v}_0\cdot(\mathbf{B}\times\mathbf{r}_0)$, arises
when Eq.~\eqref{ptran} is substituted into the $-\mathbf v_0(t) \cdot
\mathbf{P}$ term in Eq.~\eqref{eq:translation}. The second term, $-({e}/{2c})
\mathbf{r} \cdot (\mathbf{B}\times\mathbf{v}_0)$, appears on the left hand
side as a result of taking the time derivative of the exponent in
Eq.~\eqref{psitran}.  These two terms can be moved to one side of the equation
and combined using the cyclic property of the mixed product:
\begin{equation}
  -\frac{e}{2c} \mathbf{v}_0\cdot
  (\mathbf{B}\times\mathbf{r}_0)
  +\frac{e}{2c} \mathbf{r} \cdot (\mathbf{B}\times\mathbf{v}_0)
  =
  -\frac{e}{2c} \mathbf{v}_0\cdot
  \mathbf{B}\times(\mathbf{r}+\mathbf{r}_{0})
\end{equation}
which is precisely the vector potential term in $-\mathbf{v}_0\cdot
\mathbf{p}$.  This leads to the moving-frame Hamiltonian
\begin{multline}
  {H}=H_\text{d}+H_\text{Z}+H_\text{SO}+V(\mathrm{r}+\mathbf{r}_0(t),t)  \\
  +V_\text{Z}(\mathrm{r}+\mathbf{r}_0(t),t)-\mathbf{v}_0\cdot \mathbf{p}.
\end{multline}
Following Golovach et al.\cite{golovach-2004}, we now perform a canonical
transformation $H\to e^S H e^{-S} \simeq(1+S) H(1-S)=H +[S, H]$, where S is
anti-Hermitian and chosen to eliminate the original spin-orbit term.  We split
$S=S_0+S_1$ such that
\begin{align}
  [S_0,H_\text{d}]+H_\text{SO}&=0, \label{S0com}\\
  [S_1,H_\text{d}]+[S_0,H_\text{Z}]&=0,\label{S1com}
\end{align}
and choose
\begin{equation}
  S_0=\frac{im}{\hbar}\sigma^i C^{ij}r^j, \label{S0def}
\end{equation}
which satisfies Eq.~\eqref{S0com}.  This can be verified, noting that $S_{0}$
has no momentum dependence so it clearly commutes with the confining
potential, and
\begin{eqnarray}
  \left[\frac{im}{\hbar}\sigma^i C^{ij}r^j,\frac{p^2}{2m}\right]
  &=&\frac{i}{\hbar} \sigma^i C^{ij}[r^j, p^k]p^k \notag \\
  &=&-\sigma^i C^{ij} p^j.
\end{eqnarray}
With $S_0$ known, we  use Eq.~\eqref{S1com} to define $S_1$,
\begin{eqnarray}
  [H_d, S_1]
  &=\,&[S_0, H_\text{Z}]=\frac{img\mu_\text{B}}{2 \hbar}C^{ij}r^j[\sigma^i,
  \sigma^k]B^k_0 \notag \\
  &=\,&-\frac{mg\mu_\text{B} }{\hbar}C^{ij}r^j \epsilon^{ikl}B^k_0 \sigma^l
  \notag \\
  &=\,&-\frac{g\mu_\text{B}}{2}\mathbf Q \cdot (\mathbf B_0 \times \boldsymbol \sigma) \notag \\
  &=\,&\frac{g\mu_\text{B}}{2}(\mathbf B_0 \times \mathbf Q)\cdot \boldsymbol
  \sigma, \label{S1def1}
\end{eqnarray}
where $Q^i \equiv (2m/\hbar)C^{ij}r^j$.  This equation can be written in terms
of the electron's orbital states $|n\rangle$ in the dot potential,
\begin{align}
  \langle n| [H_d,S_1]|m \rangle=\,&(S_1)_{nm}(E_n-E_m) \notag \\
  =\,&\frac{g\mu_\text{B}}{2} \boldsymbol \sigma\cdot (\mathbf B_0 \times
  \langle\mathbf Q \rangle_{nm}).
\end{align}
As long as the relevant dot quantization energies are non-degenerate, $E_n
\neq E_m$, we have
\begin{align}
  (S_1)_{nm}&=\frac{g\mu_\text{B}}{2} \boldsymbol \sigma\cdot \frac{(\mathbf
    B_0 \times \langle\mathbf Q \rangle_{nm}) }{E_n-E_m} \notag \\
  &=\frac{g\mu_\text{B}}{2} \boldsymbol \sigma\cdot \mathbf{W}_{nm},
  \label{S1def}
\end{align}
where we defined
\begin{equation}
  \mathbf{W}_{nm}\equiv\frac{\mathbf B_0 \times \langle\mathbf Q \rangle_{nm} }{E_n-E_m}.
\end{equation}
Expanding the canonical transformation to first order in the spin-orbit
parameter contained in $S_{0}$ and $S_{1}$, we obtain the transformed Hamiltonian
\begin{multline}
  \bar{H}\simeq H+ [S_0, V_\text{Z}]-[S_0,\mathbf{v}_0\cdot \mathbf{p}]+[S_1,
  V] \\
  +[S_1,H_\text{Z}]-[S_1,\mathbf{v}_0\cdot \mathbf{p}]+[S_1, V_\text{Z}]
  \label{heff}
\end{multline}
Using the definition of $S_0$, the first two commutators are
\begin{align}
  \frac{img\mu_\text{B} }{2 \hbar}C^{ij}r^j[\sigma^i,\sigma^k]B_\text{N}^k
  &=-\frac{mg\mu_\text{B} }{\hbar}C^{ij} r^j(\epsilon^{ikl}B_\text{N}^k
  \sigma^l) \notag \\
  &=\frac{g\mu_\text{B} }{2}(\mathbf B_\text{N} \times \mathbf Q)\cdot
  \boldsymbol \sigma
\end{align}
and
\begin{equation}
  -\frac{im}{\hbar}\sigma^iC^{ij}[r^j,p^k]v_0^k
  =m\sigma^iC^{ij}v_0^j=\frac{1}{2}\mathbf Q_\text{v}
  \cdot \boldsymbol \sigma,
\end{equation}
respectively, with $Q_\text{v}^i \equiv 2mC^{ij}v_{0}^j$ defined analogously
to $Q^i$.  We now define the effective spin Hamiltonian by projecting onto
the orbital ground state, $H_S \equiv \langle 0 |\bar{H}|0\rangle$.  This
allows us to express the remaining commutators in the transformed Hamiltonian
using the definition of $S_1$.  The first commutator involving $S_1$ in
Eq.~\eqref{heff} is simplified by explicitly writing out the commutator
and inserting a complete set of states,
\begin{align}
  \langle 0|S_1V|0\rangle-&\langle 0|VS_1|0\rangle \notag \\
  =&\sum_{n>0}\langle 0|S_1|n \rangle \langle n |V|0\rangle -\langle 0|V|n
  \rangle \langle n |S_1|0\rangle
  \notag \\
  =&\sum_{n>0}(S_1)_{0n}V_{n0}-V_{0n}(S_1)_{n0} . \label{tech}
\end{align}
We assume $V$ to have no momentum dependence, so it commutes with $S_1$ and we
can reverse the order of the second term above.  Since $S_{1}$ is
anti-Hermitian, while $V$ is Hermitian,
\begin{align}
  \langle 0|S_1V|0\rangle-&\langle 0|VS_1|0\rangle \notag \\
  =&2\sum_{n>0}(S_1)_{0n}V_{n0} \notag \\
  =& g \mu_\text{B} \boldsymbol \sigma\cdot \sum_{n>0}\mathbf{W}_{0n}V_{n0}.
\end{align}
This technique can be used on the remaining terms in Eq.~\eqref{heff}. The
second term involving $S_{1}$ in Eq.~\eqref{heff} vanishes because
\begin{equation}
  (H_\text{Z})_{n0} = \frac{g\mu_\text{B}}{2} \boldsymbol \sigma \cdot \langle
  n|\mathbf B_0 |0\rangle=0.
\end{equation}
However, the final commutator in Eq.~\eqref{heff} contains the terms
\begin{equation}
  (S_1)_{0n}(V_\text{Z})_{n0}-(V_\text{Z})_{0n}(S_{1})_{n0}, \label{noncom}
\end{equation}
which do not commute because they contain two spin terms, but can be
treated using the spin identity
\begin{equation}
  (\boldsymbol \sigma \cdot \mathbf a)(\boldsymbol \sigma \cdot \mathbf
  b)=\mathbf a \cdot \mathbf b + i \boldsymbol \sigma \cdot (\mathbf a
  \times\mathbf b).
\end{equation}
The first term in Eq.~\eqref{noncom} becomes
\begin{equation}
  \left(\frac{g \mu_\text{B}}{2}\right)^2 \{ \mathbf{W}_{0n} \cdot
  (\mathbf{B_\text{N}})_{n0}
  +i \boldsymbol \sigma \cdot [\mathbf{W}_{0n} \times (
  \mathbf{B_\text{N}})_{n0}] \},
\end{equation}
and the second term looks quite similar, except the anti-commutator of
the cross product causes the spin dependent term to cancel with the one above,
while the spin-indepedent terms is doubled. The Hamiltonian contains
several of these spin-independent terms that can be taken as constants. The
effective spin Hamiltonian, up to a constant, can now be written simply as
\begin{equation}
  \label{eq:ham-effective}
  H_S={1\over2}\hbar\left[\boldsymbol \omega_{0}+ \boldsymbol
    \omega_1(t)
  \right] \cdot
  \boldsymbol \sigma,
\end{equation}
where
\begin{equation}
  \boldsymbol \omega_0=\frac{g \mu_\text{B}}{ \hbar} \mathbf B_0 ,
\end{equation}
is the Larmor frequency, and the time-dependent term $\omega_{1}\equiv\omega_{1}(t)$ is
\begin{align}
  \nonumber
  \boldsymbol \omega_1
  & =\frac{g \mu_\text{B}}{ \hbar}
  \left[\mathbf B_\text{N}+(\mathbf B_\text{N})_{0n} \times (\mathbf Q)_{n0} \right] +\frac{4g\mu_\text{B}}{\hbar} \mathbf{W}_{0n}V_{n0}\\
  &+\frac{1}{2 \hbar} \mathbf{Q}_\text{v} -
  \frac{2g\mu_\text{B}}{\hbar} \mathbf{W}_{0n} (\mathbf{v}_0\cdot
  \mathbf{p}_{n0}), \label{eq:omega1}
\end{align}
where the index $n$ is implicitly summed over all excited states of the dot.  If we
ignore the phonons and approximate the Overhauser fields as static, the time
dependence of the terms in the first line of Eq.~(\ref{eq:omega1}) comes only
from the position, parameterized by the known trajectory of the dot,
$\mathbf{r}_{0}(t)$.  Similarly, the time-dependence of the terms in the
second line comes from both the position $\mathbf{r}_0(t)$ and the dot
velocity $\mathbf{v}_0(t)$, [in fact, these terms are all linear in components
of $\mathbf{v}_0(t)$].  This simple spin
Hamiltonian is the key result of this derivation; it is correct to linear
order in the spin-orbit couplings.  It should be noted that including cubic spin-orbit
terms in the original Hamiltonian introduces additional terms proportional 
to $\mathbf{v}_0^{2}(t)$ and $\mathbf{v}^{3}_0(t)$ in Eq.~(\ref{eq:omega1}), but 
these terms are smaller by at least an order of
magnitude\cite{krich-2007} and the general spin
Hamiltonian form in Eq.~(\ref{eq:ham-effective}) is preserved.  

\section{Average fidelity}
\label{sec:analysis}
\subsection{General expression}

In order to analyze the implications of the additional terms in the effective
spin Hamiltonian (\ref{eq:ham-effective}), we need to take into account that
the qubit is actually formed by two electron spins.  It will be convenient to
assume that the dots' velocities are small compared to the speed of sound
$v_0\ll s\sim 5\times 10^3\,$m/s.  Then the phonon effects should decouple
from the dots' motion and can be approximated as contributing to the same
``intrinsic'' decoherence times $T_1$, $T_2$ as one would have without the
motion.  The effect of such decoherence terms on dynamical decoupling has been
considered in detail in Ref.~\onlinecite{pryadko-quiroz-2009}; in the
following we assume that these decoherence times
are large compared to the characteristic period $T$ of the dots' motion and
therefore can be ignored.

In the absence of phonons, and approximating the Overhauser field as
classical, the time evolution of the two-spin wavefunction with $N=4$
components can be characterized by the unitary matrix $U(t)$.  The qubit
subspace $\mathcal Q$ is formed by the $m_z=0$ component of the two-spin
wavefunction; it has $M=2$ dimensions.  The standard assumption is that, at
the beginning of the experiment, the two spins are initiated in a pure state
$\ket\psi$ which belongs to the qubit subspace, $\ket\psi\in\mathcal{Q}$.
Therefore, when computing the average fidelity, we need to average only over
the original wavefunctions in $\mathcal{Q}$.

More generally, consider an $M$-dimensional subspace $\mathcal{Q}$ of an
$N$-dimensional Hilbert space $\mathcal{H}$.  Let us introduce an $N\times M$
matrix $T$ whose columns are formed by the components of orthonormal vectors
forming a basis of $\mathcal{Q}$.  Then, the components of an arbitrary
wavefunction $\ket \psi\in\mathcal{Q}$ are a linear combination of the columns
of $T$; namely $\psi=T\varphi$, where $\varphi$ is an $M$-dimensional column
vector, $\|\varphi\|=1$.  The corresponding density matrix can be written in
this basis as $\rho_0\equiv T \varphi\,\varphi^\dagger T^\dagger$.  The
fidelity corresponding to the evolution matrix $U$ is
\begin{equation}
  F=\tr (\rho_0 U \rho_0 U^{\dagger})=(\varphi^\dagger W\varphi)(\varphi^\dagger W^\dagger\varphi),
\end{equation}
where $W=T^{\dagger}UT$ can be thought of as the projection of $U$ onto the
subspace $\mathcal{Q}$.  The average fidelity in the subspace can now be
calculated using the averaging identities for components $\varphi_i$ of the
normalized wavefunction $\ket\varphi$
\begin{align}
\langle \varphi_i \varphi_j^* \rangle &= \delta_{ij} / M, \\
\langle \varphi_i \varphi_j^*\varphi_k \varphi_l^* \rangle &=
\frac{\delta_{ij}\delta_{kl}+\delta_{il}\delta_{jk}}{M^2+M}.
\end{align}
This leads to the
average fidelity
\begin{equation}
  \langle F \rangle = \frac{|\tr W|^2+\tr(W W^\dagger)}{M^2+M}. \label{eq:av-fid}
\end{equation}

For the special case of the qubit formed by the singlet and $m=0$ triplet
states of two spins, assuming no interdot tunneling, we can take the net
evolution matrix as the Kronecker product of evolution matrices corresponding to
the two qubits, $U=U_1\otimes U_2$.  Further, it will be convenient to
decompose each single-spin matrix in the interaction representation with
respect to the precession in the net effective magnetic field along the
$z$-axis,
\begin{equation}
  U_i=U_{0i} S_i,\quad U_{0i}\equiv e^{-i\sigma^z\,\varphi_i(t)/2},
  \label{eq:unitary-decomposition}
\end{equation}
where
\begin{equation}
  \varphi_i(t)
  \equiv \omega_0t +\int_0^t dt'\, {\omega}^z_{1i}(t'),
  \label{eq:unperturbed}
\end{equation}
$\omega_0$ is the Larmor frequency, $\omega_{1i}^z(t)$, $i=1,2$
[cf.~Eq.~(\ref{eq:ham-effective})] are the two dot's effective perturbing
fields in the $z$-direction, and the matrices
\begin{equation}
  \label{eq:unitary-slow}
  S_i\equiv e^{-i\gamma_i-i{\boldsymbol\phi}_i\cdot{\boldsymbol\sigma}/2}, \;\,i=1,2,
\end{equation}
are parametrized as rotations by angle $\phi_i\equiv |{\boldsymbol\phi}_i|$
around the unit vectors $\hat{\boldsymbol\phi}_i$, with extra phases
$\gamma_i$.  These rotations come entirely from transverse, $\mu=x,y$,
components of ${\boldsymbol\omega}^\mu_{1i}$ in the rotating frame, largely due
to the Larmor frequency.  Since the Larmor frequency is large, the additional
rotation angles are expected to be small; we expand the average
fidelity~(\ref{eq:av-fid}) to quadratic order in components of
${\boldsymbol\phi}_i$,
\begin{eqnarray}
  \label{eq:av-infid-expanded}
  \langle F\rangle &=& 1-f_0-f_1-f_2^z-f_2^\perp+\ldots,
\end{eqnarray}
with the infidelity terms
\begin{eqnarray}
  f_0&=&{2\over 3}\sin^2
  (\Delta\varphi/2),\\
  f_1&=&{1\over 3}(\phi_2^z-\phi_1^z)\sin (\Delta \varphi),\\
  f_2^z&=&{1\over6}
  (\phi_2^z-\phi_1^z)^2\cos(\Delta \varphi),\\
  f_2^\perp&=& {2+\cos (\Delta \varphi)\over 12}[(\phi_1^\perp
  )^2+(\phi_2^\perp )^2].
\end{eqnarray}
Note that, as expected, the fidelity only depends on the differences
$\Delta\varphi\equiv \varphi_2-\varphi_1$ and $\phi_2^z-\phi_1^z$ of the two
precession angles around the $z$-axis.  To the same quadratic accuracy in the
small angles, we can also write
\begin{equation}
  \label{eq:infid-z-combined}
  f_0+f_1+f_2^z={1\over 3}[1-\cos(\Delta\varphi+\phi_2^z-\phi_1^z)],
\end{equation}
which only depends on the total phase difference, and is exact in the
limit $\phi_i^\perp=0$.

\subsection{Rotating-frame approximation}

We now return to the analysis of the single-spin
Hamiltonian~(\ref{eq:ham-effective}).  The Larmor frequency,
$\omega_0\agt 4\times 10^{9}$\,rad/s, is the dominant term, that is,
$\omega_1\ll \omega_0$.  Given that we have excluded the phonons, we
can also assume that the dot trajectory is such that the time
dependence in $\omega_1(t)$ is slow on the scale of $\omega_0$.  This
implies that we can use the interaction representation Eq.~(\ref{eq:unitary-decomposition}), 
where $S(t)$ is the slow part of the unitary evolution operator; it obeys
the equation
\begin{equation}
  \label{eq:evol-slow}
  i\hbar \dot S= H_\mathrm{int}(t) S,\quad S(0)=\openone,
\end{equation}
where we temporarily omit the dot index, and
\begin{eqnarray}\nonumber
\lefteqn{{1\over \hbar} H_\mathrm{int}(t)\equiv
{1\over2}\boldsymbol\omega_1(t)
\cdot
U_0^\dagger(t)\boldsymbol \sigma U_0(t)}& & \\
&=&{1\over2}\Bigl\{\sigma^x[\omega_1^x(t) \cos \varphi(t)
+\omega_1^y(t) \sin \varphi(t)]\nonumber\\ & &\;\; +
\sigma^y[\omega_1^y(t) \cos \varphi(t)-\omega_1^x(t) \sin
\varphi(t)]
\Bigr\}\;\;\;\quad
\label{eq:pert}
\end{eqnarray}
is the perturbing Hamiltonian in the interaction representation.

The remaining analysis is performed perturbatively with the Magnus
(cumulant) expansion,
\begin{equation}
  S(t)=\exp ({\cal C}^{(1)}+{\cal C}^{(2)}+\ldots),
\end{equation}
where the first two cumulants are
\begin{eqnarray}
  {\cal C}^{(1)}(t)&=&-\frac{i}{\hbar}\int_{0}^{t}dt_{1}
  H_\text{int}(t_{1}), \label{eq:U1}\\
  {\cal C}^{(2)}(t)&=&-{1\over2\hbar^2}\int_{0<t_{1}<t_{2}<t} \hskip-0.4in
  \,dt_{1} \,dt_{2}\,[ H_\mathrm{int}(t_{2}), H_\mathrm{int}(t_{1})].
  \label{eq:U2}
\end{eqnarray}
We perform the integration explicitly to leading order in
$\omega_1/\omega_0$, also assuming that the time-dependence of $\omega_1$ is
slow on the scale of $\omega_0$:
\begin{eqnarray}
    i\mathcal{C}^{(1)}(t)&=&{1\over 2\omega_0}\Bigl\{\sigma^x
    [a\sin\varphi(t)+ b_0-b\cos \varphi(t) ]\quad\nonumber \\
    \label{eq:C1}
    & & \hskip-0.3in  +\sigma^y[b\sin\varphi(t)-
    a_0+a\cos \varphi(t)]\Bigr\},\\ \label{eq:C2}
    i\mathcal{C}^{(2)}(t)&=&{1\over 4\omega_0}\sigma^z \int_0^t dt_1\,
    [a^2(t_1)+b^2(t_1)] .
\end{eqnarray}
Here we denoted $a\equiv \omega_1^x(t)$, $b\equiv \omega_1^y(t)$, and $a_0$,
$b_0$ the corresponding values at $t=0$ [$\varphi(0)\equiv 0$ by definition].

Eq.~(\ref{eq:av-infid-expanded}) gives the expression for the qubit infidelity
\begin{eqnarray}
  \label{eq:Ifinal}
\lefteqn{ 1-\langle F\rangle =f_z+f_1^\perp+f_2^\perp,}\quad & & \\
\label{eq:Iz}
f^z&=&{1\over3}[1-\cos\left(\Delta\varphi+\delta_2-\delta_1\right)],
\\  \label{eq:Iperp}
 f_i^\perp&=&{2+\cos(\Delta\varphi)\over
  12\omega_0^2}\left[ (B_i-b_{0i})^2+(A_i-a_{0i})^2\right],\quad
\end{eqnarray}
where the index $i=1,2$ refers to the two spins, $A_i\equiv a_i \cos\varphi(t)
+b_i \sin\varphi(t)$ and $B_i\equiv b_i \cos\varphi(t) -a_i \sin\varphi(t)$
are the rotated components of the transverse angular-velocity vectors
$(\omega_1^x,\omega_1^y)$ for the corresponding spins, and the additional
phases $\delta_i$ are given by the integrals.
\begin{equation}
  \label{eq:addl-phase}
  \delta_i\equiv  \int_0^t{dt'\over
    2\omega_0}\,[\omega_{1i}^\perp (t')]^2.
\end{equation}
One immediately recognizes the additional phases in
Eqs.~(\ref{eq:Iz}) and (\ref{eq:addl-phase}) as the effect of level
repulsion, or equivalently as the gap for the spins, driven
adiabatically by the total instantaneous magnetic field $\propto
[(\omega_0+\omega_{1i}^z)^2+(\omega_{1i}^\perp)^2]^{1/2}$.  Then,
Eq.~(\ref{eq:Iperp}) can be interpreted as the effect of the basis
change between the original quantization $z$-axis and the direction of
the instantaneous magnetic field; it is similar in nature to the
initial decoherence associated with any periodic decoupling sequence, see, e.g. Ref.~\onlinecite{pryadko-sengupta-kinetics-2006}.

The terms of the next order in $1/\omega_0$ expansion, omitted in
Eqs.~(\ref{eq:C1}) and (\ref{eq:C2}), include the trivial correction to
Eq.~(\ref{eq:C2}) $\propto \int
dt\,\omega_1^z[\omega_1^\perp]^2/\omega_0^2$, as well as the geometrical
phase $\propto \int dt'\,W[\omega_{1}^x,\omega_1^y]$, where
$W[x,y]\equiv W[x(t),y(t)]\equiv x(t)y'(t)-x'(t) y(t)$ is the Wronskian.

Since $\Delta\varphi=\int_0^t
dt'\,[\omega_{12}^z(t')-\omega_{11}^z(t')]$, the term $f^z$ does not
contain any rapid oscillations at the Larmor precession frequency
$\omega_0$, while the terms $f_i^\perp$ can be averaged over the
period of Larmor precession by replacing $A_i^2 +B_i^2$ with
$(\omega_{1i}^\perp)^2$ and dropping all of the terms linear in $A_i$,
$B_i$,
\begin{equation}
  \label{eq:IF-perp-av0}
  {\overline f}_i^\perp={2+\cos(\phi_2^z-\phi_1^z)\over
    3\omega_0^2}\left\{[\omega_{1i}^\perp(t)]^2
    +[\omega_{1i}^\perp(0)]^2\right\}.
\end{equation}

\subsection{Sequences}
Overall, the dynamical decoupling should be designed to null the difference
between the accumulated phases $\varphi_i+\delta_i$ of the two spins which
suppresses the main contribution to infidelity, see Eq.~(\ref{eq:Iz}).  For a
static Overhauser field, this can be achieved just by ensuring that each spin
spends the same amount of time at each position, e.g., via the solid adiabatic
trajectory in Fig.~[\ref{fig:protocols}].  This is also sufficient to suppress the effect of
the velocity-dependent terms in the second line of Eq.~(\ref{eq:omega1}).  A
more complicated set of dot rotation involving motion in both directions,
e.g., see the dashed trajectory of Fig.~[\ref{fig:protocols}], are required to suppress a time-dependent
Overhauser field.

To analyze the effect of a sequence of $\pi$-rotations in the presence of a
time and position-dependent Overhauser field, consider its Fourier expansion
at a position on the trajectory parametrized by the rotation angle $\theta$,
\begin{equation}
  B^z(\theta,t)=A_0(t)+\sum_m A_m(t) \cos m\theta +B_m(t) \sin m\theta.
  \label{eq:field-expansion}
\end{equation}
Only the difference between the fields corresponding to the two dots (located
at $\theta$ and $\theta+\pi$) is relevant for the infidelity
Eq.~(\ref{eq:Iz}).  This leaves only the terms with $m$ odd in the Fourier
expansion~(\ref{eq:field-expansion}).

For a term with $\cos m\theta$ (an even function of $\theta$) a sequence of
$\pi$ rotations acts the same way, independent of the direction.  It is easy to
check that with an equidistant sequence of rotations centered at $T_0/2$,
$3T_0/2$, \ldots, $(2s+1)T_0/2$, where the number of rotations $s$ is even;
any time-independent and linear in $t$ contributions to $A_m(t)$ are
suppressed, but a quadratic term would generally remain.  Unlike the usual
dynamical decoupling problem\cite{khodjasteh-Lidar-2005,Uhrig-2007}, it is not
generally possible to suppress the quadratic term of $A_m(t)$.

The rotation direction starts to matter for a term with $\sin m\theta$ which
is an odd function of $\theta$.  Here a sequence of $\pi$ rotations in the
same direction picks up a sum of contributions from consecutive time intervals
with alternating signs, suppressing the time-independent contribution to
$B_m(t)$ but not the linear contribution.  As an alternative prescription, a
symmetrized sequence of two forward rotations by angle $\pi$, followed by two
rotations in the opposite direction can be used to suppress the effect of the
linear in $t$ term in $B_m(t)$.  Generally, it is possible to design more
complicated sequences analogous to concatenated\cite{khodjasteh-Lidar-2005} or
Uhrig's\cite{Uhrig-2007} sequences to suppress the effect of any fixed-degree
polynomial in time $B_m(t)$.  However, we do not expect this to be useful
since the quadratic time contribution of $A_m(t)$ would still remain.

\section{Simulations}
\label{sec:simulations}
We corroborate our analytical results by simulating the two-spin unitary
evolution with the effective Hamiltonian~(\ref{eq:ham-effective}).
Specifically, we parametrize the dot trajectory by the rotation angle
$\theta=\theta_1(t)$; the other dot is assumed to have the symmetric position,
$\theta_2(t)=\theta_1(t)+\pi$ [see Fig.~\ref{fig:protocols} for samples of
actual trajectories.]  The position-dependent terms in the first line of
Eq.~(\ref{eq:omega1}) are simulated in terms of a three-component correlated
magnetic field $\mathbf{B}(\theta)$ drawn from the Gaussian distribution with
zero average and the correlation function $\langle B_\mu
(\theta)B_\nu(\theta')\rangle=\sigma_\mu^2
\delta_{\mu\nu}\vartheta(\theta-\theta')$ (no implicit summation in
$\mu,\nu=x,y,z$), where
$$
\vartheta(\theta)\equiv \sum_{m=-\infty}^\infty e^{-(\theta-2\pi
  m)^2/\ell^2}
$$
is an infinite sum of Gaussian functions (which can also be represented in
terms of the Jacobi theta function).  These are obtained by applying a
Gaussian filter to a discrete set of uncorrelated random numbers drawn from
the Gaussian distribution, and using the standard cubic spline interpolation
with the result.  To simulate the components of time-dependent magnetic field
${B}(\mathbf{r},t)$, we used explicit order-$r$ spline interpolation between
several such angle-dependent functions, where $r=1,2$. 

For all simulations, we chose the time units corresponding to the
Larmor precession period $\tau_0\equiv 2\pi/\omega_0$ and the correlation
time of the Overhauser field $4\cdot10^4\tau_0$ with each component of its 
r.m.s.\ value corresponding to rotation frequency
$\langle|\boldsymbol{\omega}_1|^2\rangle=0.025/\tau_0^2$.  The adiabatic trajectories
of the dots were simulated using a sum of appropriately shifted hyperbolic tangents, scaled
so that the dot is in motion during approximately half of the protocol.  The leading velocity-dependent
term in Eq.~(\ref{eq:omega1}) was simulated using the corresponding derivatives and the
parameter $Q_v/\hbar=0.075/\tau_0$, assuming equal contributions from the Rashba and 
Dresselhaus parameters.

\begin{figure}[htbp]
  \centering
  \includegraphics[width = 0.9\columnwidth]{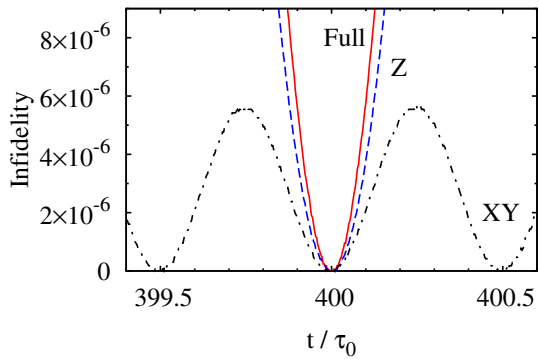}
  \caption{Simulated qubit infidelity $1-\langle F\rangle$
    [Eq.~(\ref{eq:av-fid})] in the vicinity of the first full rotation period
    of the double-dot qubit at $t=T$.  Position-dependent magnetic field
    $B_\mu(\theta)$ is assumed static, and the rotation period $T$ is chosen
    commensurate with the Larmor frequency, $\omega_0 T/2\pi=400$.
    Dashed line: only $B_z(\theta)$ is included; the infidelity minimum is
    exactly at $t=T$, in agreement with Eq.~(\ref{eq:Iperp}) which is exact in
    this situation.  Dotted line: only the transverse components
    $B_\mu(\theta)$, $\mu=x,y$ are included.  The infidelity minimum is
    slightly off the commensurate time $t=T$ due to the terms not included in
    Eq.~(\ref{eq:Ifinal}).  All
    three components of the field $B_\mu(\theta)$ are included for the red solid
    line.}
  \label{fig:static}
\end{figure}

\begin{figure}[htbp]
  \centering
  \includegraphics[width = 0.9\columnwidth]{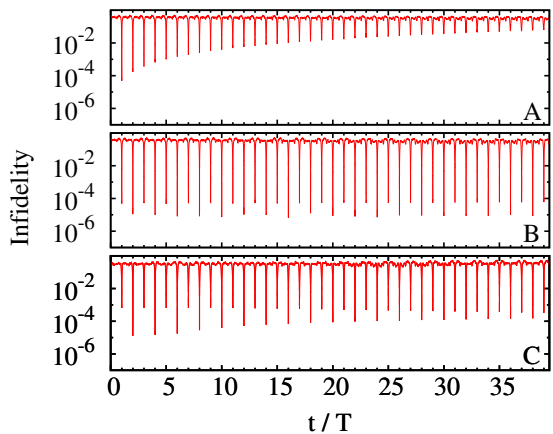}
  \caption{Simulation results for fidelity measured at the end of each
    cycle, for a linear time-dependence of $B_{1i}^\mu$ with (a) forward, (b) forward-back dot rotations
    [cf.~Fig.~\ref{fig:protocols}], as well as (c) forward-back for quadratic time-dependence.}
  \label{fig:multicycle}
\end{figure}

For the case of the static Overhauser field, see Fig.~\ref{fig:static}, we see 
that the average infidelity is dominated by the contribution from the $z$-component
of the field, and that the infidelity nearly vanishes at the end of the spatial exchange
protocol.  For a linearly time-interpolated Overhauser field,
the infidelity increases over several cycles of the single-direction $\pi$ pulses,
Fig.~\ref{fig:multicycle}a, but maintains a low value $\sim10^{-5}$ after alternating between a sequence of two forward rotations, followed by two rotations in the opposite direction, Fig.~\ref{fig:multicycle}b.  However, for the quadratic interpolation, Fig.~\ref{fig:multicycle}c, the infidelity gradually increases even for the alternating 
protocol, though it stays below $10^{-4}$ for several cycles.

\section{Possible Experimental Setup}
\label{sec:experimental-proposal}
Aspects of our proposal, such as the precise construction of few-electron
quantum dots in III-V semiconductor heterostructures have already been
demonstrated in experiments. \cite{petta-2005, reilly-2008, vanWeperen-2011,
  laird-2010} However, the precise adiabatic rotation required in our proposal
may be quite difficult to accomplish experimentally.  We now discuss possible
design implementations for an experimental realization, as well as physical
constraints.

As discussed above, our proposal is suitable in materials with relatively weak
spin-orbit coupling such as GaAs/AlGaAs heterostructures.  We believe the
electrons in a 2DEG could be confined in the radial direction by creating a
circular depletion layer by placing electrostatic gates with a constant
voltage in the center and outer edge of the circle as sketched in
Figure~\ref{fig:setup}.  Since we wish to suppress the tunneling, the normal
interdot spacing should be much larger than the dot size, $a$, so we assume
that the circular trajectory has a radius $r_0 \sim 15a \sim 1$~{\textmu}m.
Confinement and rotation in the angular direction could be accomplished by
placing appropriately chosen time-dependent voltages on the "wedge-gates" on
both sides of the electron (Fig.~\ref{fig:setup}).  Several of these wedges
will be needed to accomplish the smooth and adiabatic trajectory needed.  The
combined use of wedge and circular gates may require gates on both the top and
bottom of the sample.  The typical confining potential is approximated by $U
\sim m\omega_\text{d}^2a^2$, which only requires reasonable gate voltages
on the order of 100 mV.

The averaging of the hyperfine interaction is only valid in the quasi-static
approximation of the Overhauser field.  In general, the hyperfine interaction
between the electron and the nuclei leads to a Knight shift.  However, in the
presence of the magnetic field, fluctuating corrections to the quasi-static
approximation are inversely proportional to $B_{0}$ and can be
neglected\cite{taylor-2007}.  Thus, the most significant effect in this case
is due to the dipole-dipole interaction of nearest neighboring nuclei which
requires that $T \ll t_{\rm nuc} \sim 10^{-4}$s\cite{merkulov-2002}.  This
places a lower bound on the velocity, while there is also an upper bound
necessary to ensure that the Lorentz force from the dot rotation only deforms
the actual path of the electron by a distance much smaller than the
correlation length of the Overhauser field.  This results in the restriction,
$10^{-1} \ll v \ll 10^{5}$ in m/s.  This rather lenient constraint is due to
the assumption that confinement potential be large compared to the other
potentials in our system.  This allows us to neglect trajectory deviations
from perturbations such as charge noise.  In our estimates we also assumed $v$
to be small compared to the speed of sound, $s\sim 5\times 10^3$m/s.  With
$a/r_0\sim15$ and $v\sim10\,$m/s we obtain the rotation period $T\sim
1$~{\textmu}s, which easily satisfies the above conditions and, according to our
simulations, should result in the infidelities lower than $10^{-4}$ for
significant time-scales.  The rotation period could potentially be decreased
to allow more operations to be performed before the states decohere.

\section{Conclusion}
\label{sec:conclusion}
 We analyzed the real-space exchange of quantum dots as a
possible substitute for the tunneling exchange.  Ideally, exchange eliminates
the hyperfine dephasing from the Overhauser field parallel to the applied
field, leaving only the smaller effects from the in-plane field.  The
real-space exchange accomplishes the same suppression of the hyperfine
interaction, but avoids the problematic sensitivity to charge noise present in
exchange via tunneling.  While spatial exchange does introduce additional
effects such as spin-orbit coupling, simple tricks like using pairs of $\pi$
rotations in alternating directions can be used to suppress these so that 
the decoherence is still dominated by the hyperfine interaction.  In
particular, this field only enters as the small ratio of the average in-plane
Overhauser field to the externally applied field.  Perhaps the simplest way to
suppress the hyperfine interaction in this approach is to reduce this ratio by
increasing the externally applied field.

In addition, this spatial exchange is compatible with some of the methods
already being attempted such as nuclear polarization via pumping.  If the
hyperfine interaction can be further suppressed, the next largest
contribution from our spatial exchange approach would come from the disorder
of the sample or the electron-phonon coupling.  Our analysis of this spatial
exchange also remains valid in systems that use additional quantum
dots,\cite{laird-2010, vanWeperen-2011} with universal quantum operations in
mind, as long as each operation is applied to only two dots at a time.

While the movement of quantum dots requires the precise control of the
confining potential, which may be difficult to realize experimentally, our
analysis shows that the construction of such a system is viable.  With
realistic parameter values from current experiments, our analysis produces
infidelities smaller than $10^{-4}$ after ten decoupling cycles.  This setup could also be a productive step towards the
experimental realization of more complicated exchange systems, with many more
interesting applications.

\section{Acknowledgements}  This work was
supported in part by the U.S. Army Research Office under Grant No.
W911NF-11-1-0027 (LPP, DD), and by the NSF under Grants DMR-0748925 (KS, DD) and
1018935 (LPP, DD).

%

\end{document}